\documentclass[12pt]{article}
\usepackage{epsfig}
\usepackage{multicol}
\usepackage{graphicx, slashed, amsmath, amsfonts, amssymb}
\usepackage[top=1in, bottom=1in, left=1in, right=1in]{geometry}
\def\bc{\begin{center}}
\def\ec{\end{center}}
\def\be{\begin{equation}}
\def\ee{\end{equation}}
\def\ba{\begin{array}}
\def\ea{\end{array}}
\def\beqn{\begin{eqnarray}}
\def\eeqn{\end{eqnarray}}
\makeatletter
\newcommand*{\rom}[1]{\expandafter\@slowromancap\romannumeral #1@}
\makeatother
 
\begin{document}
\title{Minimal texture of quark mass matrices and precision CKM measurements}
\author{Shivali Kaundal, Aakriti Bagai, Gulsheen Ahuja*, Manmohan Gupta$^{+}$\\
{\it Department of Physics, Centre of Advanced Study,}\\
{\it Panjab University, Chandigarh, India.}\\
*gulsheen@pu.ac.in, $^{+}$mmgupta@pu.ac.in
{}}\maketitle

\begin{abstract}
We have carried out an extensive analysis of all possible minimal texture quark mass matrices implying 169 texture 6 zero combinations. One finds that all these combinations are ruled out, a good number of these analytically, the other possibilities are excluded by the present quark mixing data. Interestingly, even if in future, there are changes in the ranges of the light quark masses, these conclusions remain valid. 
\end{abstract}

Over the last couple of decades, noticable progress has been made in measuring the quark mixing parameters. Most of the Cabibbo Kobayashi Maskawa (CKM) \cite{cab,km} parameters are now known within an accuracy of around 5$\%$, this can be considered as being at the level of `precision measurements' in the context of CKM phenomenology. Similarly, good deal of progress has been made in the measurement of the quark masses, in particular, the light quark masses, $m_u$, $m_d$ and $m_{s}$ have registered remarkable progress in their measurements in the last decade. This has been possible due to the lattice simulations providing the most reliable determination of the strange quark mass and of the average of the up and down quark masses, as emphasized by Flavour Lattice Averaging Group (FLAG) \cite{flag}. In view of the relationship of CKM matrix with the mass matrices, precision measurements of CKM parameters would undoubtedly have implications for the quark mass matrices. Similarly, considerable narrowing of the ranges of the light quark masses would allow us to determine the nature and structure of the quark mass matrices which are compatible with the CKM phenomenology.

It is well known that the mass matrices, having their origin in the Higg's fermion couplings, are arbitrary in the Standard Model (SM), therefore the number of free parameters available with a general mass matrix is 36 which is much larger than the number of physical observables, e.g., in the quark sector, the 10 observables include 6 quark masses, 3 mixing angles and 1 CP violating phase. Therefore, to develop viable phenomenological fermion mass matrices one has to limit the number of free parameters in these matrices. It may be noted that in the SM as well as its extensions, wherein the right handed quarks are singlets, without loss of generality, one can always consider the mass matrices to be hermitian. In this context, the idea of texture zero mass matrices \cite{fxrevart}-\cite{ourrevart2} consisting of finding the phenomenological quark mass matrices which are in tune with the low energy data, i.e., observables like quark masses, quark mixing angles, angles of the unitarity triangle in the quark sector, etc., has proved to be quite successful in explaining the fermion mixing data. A particular square matrix is considered to be texture `n' zero if the sum of the number of diagonal zeros and half the number of the symmetrically placed off diagonal zeros is n.

In the quark sector, the concept of texture zeros was introduced implicitly by Weinberg \cite{wein} and explicitly by Fritzsch \cite{friori1,friori2}, the original Fritzsch \emph{ans$\ddot{a}$tz} being given by 
 \begin{equation}
    M_{U}=\begin{pmatrix}
      0 & A_{U} & 0 \\
      A^{*}_{U} & 0 & B_{U} \\
      0 & B_{U}^{*} & C_{U} \\
    \end{pmatrix} ,\quad M_{D}=\begin{pmatrix}
      0 & A_{D} & 0 \\
      A^{*}_{D} & 0 & B_{D} \\
      0 & B_{D}^{*} & C_{D} \\
    \end{pmatrix},
\label{mumd}
\end{equation}
where $M_{U}$ and $M_{D}$ correspond to the mass matrices in the up (U)  and down (D) sector with complex off diagonal elements, i.e., $A_{i}=|A_{i}|e^{\text{i$\alpha $}}$ and $B_{i}=|B_{i}|e^{\text{i$\beta $}}$, where i = \emph{U,D}, whereas $C_{i}$ is the real element of the matrix. Using the above definition of texture zero mass matrices, each of the above matrix is said to be texture 3 zero type, together these are referred as texture 6 zero quark mass matrices. The above mentioned original Fritzsch \emph{ans$\ddot{a}$tz} as well as several other texture 5 and 4 zero versions have been examined in references \cite{tsmmrrr}-\cite{tsmmgp4}. In particular, references \cite{tsmmrrr} and \cite{tsmmneelu} have discussed texture 6 zero quark mass matrices and have arrived at the conclusion that these look to be incompatible with the quark mixing data. However, in these references, a detailed and comprehensive analysis indicating how and to what extent these matrices are ruled out has not been discussed. In particular, neither the references consider all possible texture 6 zero quark mass matrices nor do these relate the various possibilities through permutation symmetry which has gained significance in the context of quark-lepton symmetry due to which the emphasis has now shifted to formulate the texture structure of fermion mass matrices incorporating permutation symmetry and Abelian symmetries.

In the absence of any firm theoretical foundation of choosing a particular texture for the mass matrices, it becomes interesting to analyse all possible texture structures of the mass matrices for checking their viability with the present refined data. It may be noted that the maximum number of texture zeros which can be introduced in the quark mass matrices is 3 in each sector, resulting in minimal number of parameters or elements of the mass matrices. In view of this, we refer texture 6 zero quark mass matrices as minimal texture of quark mass matrices. It is interesting to note that not only the matrices mentioned in equation (\ref{mumd}) correspond to texture 3 zero mass matrices each, but along with these, there are several other possible structures which can be considered to be texture 3 zero ones. The purpose of the present work is to first enumerate all possible minimal texture quark mass matrices, i.e., texture 6 zero mass matrices and to relate these possibilities using permutation symmetry. As a next step, we have examined in a detailed and comprehensive manner the viability of all these possible mass matrices keeping in mind the improvements in the measurements of light quark masses $m_u$, $m_d$ and $m_{s}$ as well as `precision measurements' of the CKM parameters.

One can check that the total number of structures for a texture `n' zero mass matrix comes out to be 
\begin{equation}
^6C_n = \frac{6!}{n!(6-n)!},
\end{equation}
where 6 is the number of ways to enter zeros in the mass matrices. Using this, for n=3 one can arrive at the following 20 possible structures (${S_0}$ to $S_{19})$ for texture 3 zero mass matrices:

\begin{list}{(i)}
	\item Placing all three zeros along diagonal positions:\\	
\begin{center}
$S_{0} =	\begin{pmatrix}
		0 & \times & \times \\ 
		\times & 0 & \times \\ 
		\times & \times & 0
	\end{pmatrix},$\\ 
\end{center}	
where $\times$'s represent the non-vanishing entries.
\end{list}

\begin{list}{(ii)}
 \item Placing two zeros along diagonal positions:\\
 \begin{center}
	$S_{1}= \begin{pmatrix}
	0	& \times & 0 \\ 
	\times	&0  &  \times\\ 
	0	&\times  & \times
	\end{pmatrix},\: S_{2}= \begin{pmatrix}
	0	& 0 & \times \\ 
	0	&\times  &  \times\\ 
	\times	&\times  & 0
\end{pmatrix},\:
S_{3}= \begin{pmatrix}
0	& \times & \times \\ 
\times	& 0  & 0\\ 
\times	& 0  & \times
\end{pmatrix},$\\
$S_{4}= \begin{pmatrix}
\times	& \times & 0 \\ 
\times	& 0  & \times\\ 
0	& \times  & 0
\end{pmatrix}\:S_{5}= \begin{pmatrix}
0	& \times & \times \\ 
\times	&\times  & 0\\ 
\times	& 0  & 0
\end{pmatrix},\: S_{6}= \begin{pmatrix}
\times	& 0 & \times \\ 
0	& 0  & \times\\ 
\times	& \times  & 0
\end{pmatrix},$

$S_{7}= \begin{pmatrix}
0	& 0 & \times \\ 
0	& 0  & \times\\ 
\times	& \times & \times
\end{pmatrix},\: S_{8}= \begin{pmatrix}
0	& \times & 0 \\ 
\times	&\times  &\times\\ 
0	& \times  & 0
\end{pmatrix}, \:S_{9}= \begin{pmatrix}
\times	& \times & \times \\ 
\times	& 0  & 0\\ 
\times	& 0  & 0
\end{pmatrix}.$
\end{center}
\end{list}
\begin{list}{(iii)}
	\item Placing one zero along diagonal position:\\
	\begin{center}
		$ S_{10}= \begin{pmatrix}
			0	& \times & 0 \\ 
			\times	& \times  & 0\\ 
			0	& 0  & \times
		\end{pmatrix},\: S_{11}= \begin{pmatrix}
			0	& 0 & \times \\ 
			0	& \times &  0\\ 
			\times	& 0  & \times
		\end{pmatrix},\:
		S_{12}= \begin{pmatrix}
		\times	& \times & 0 \\ 
			\times	&0  &  0\\ 
			0	& 0  & \times
		\end{pmatrix},$\\

	$	S_{13}= \begin{pmatrix}
			\times	& 0 & 0 \\ 
			0	& \times  &  \times\\ 
			0	& \times  & 0
		\end{pmatrix},\: S_{14}= \begin{pmatrix}
			\times	& 0 & \times \\ 
			0	& \times  &  0\\ 
			\times	& 0  & 0
		\end{pmatrix},\:
		S_{15}= \begin{pmatrix}
			\times	& 0 & 0 \\ 
			0	& 0  & \times\\ 
			0	& \times & \times
		\end{pmatrix},$\\
	$S_{16}= \begin{pmatrix}
		\times	& \times & 0 \\ 
		\times	& \times  &  0\\ 
		0	& 0  & 0
	\end{pmatrix},\: S_{17}= \begin{pmatrix}
		\times	& 0 & \times \\ 
		0	& 0  &  0 \\ 
		\times	& 0  &\times
	\end{pmatrix},\:
	 S_{18}= \begin{pmatrix}
    	0		& 0 & 0 \\ 
		0	& \times  & \times\\ 
		0	& \times  & \times
	\end{pmatrix}.$
	\end{center}
	\end{list}
\begin{list}{(iv)}
	\item Placing all zeros in off diagonal positions:\\
	\begin{center}
	$S_{19}= \begin{pmatrix}
			\times	& 0 & 0 \\ 
			0	& \times  &  0\\ 
			0	& 0  & \times
		\end{pmatrix}.$
	\end{center}
    \end{list}    
    
In general, one has the freedom to consider the mass matrices in the up and down sectors, i.e., $M_{U}$ and $M_{D}$ to be either  of the above listed 20 patterns, resulting into 400 combinations corresponding to texture 6 zero mass matrices. However, since these matrices have to yield physical quark masses as their eigenvalues, therefore, the trace as well as determinant of these should be non zero, i.e., 
\begin{equation}
{\rm Trace} M_{U,D} \neq 0~~~~{\rm and}~~~{\rm Det}M_{U,D} \neq 0.
\end{equation}
Imposing these constraints, one can immediately see that either the trace or the determinant of the structures $S_0$, $S_7$, $S_8$, $S_{9}$, $S_{16}$, $S_{17}$ and $S_{18}$ vanishes and hence out of 20 possible patterns, we are left with 13 structures to be considered either as $M_{U}$ or $M_{D}$, leading to 169 possible texture 6 zero combinations.

It may be noted that structure $S_1$ is in fact the Fritzsch \emph{ans$\ddot{a}$tz} mentioned in equation (\ref{mumd}). Interestingly, one finds that the structures $S_1$, $S_2$, $S_3$, $S_4$, $S_{5}$ and $S_{6}$ are related as
\be
S_j = p_j^T S_1 p_j, ~~~~~~~~(j=1-6)
\ee
where $p_j$ are the following 6 permutation matrices
\begin{equation}
p_1=\left(
\begin{array}{ccc}
	1 & 0 & 0 \\
	0 & 1 & 0 \\
	0 & 0 & 1
\end{array}
\right) ,\: p_2=\left(
\begin{array}{ccc}
	1 & 0 & 0 \\
	0 & 0 & 1 \\
	0 & 1 & 0
\end{array}
\right),\: p_3=\left(
\begin{array}{ccc}
	0 & 1 & 0 \\
	1 & 0 & 0 \\
	0 & 0 & 1
\end{array}
\right),
\end{equation}	\\
\begin{equation}
	p_4=\left(
	\begin{array}{ccc}
		0 & 0 & 1 \\
		0 & 1 & 0 \\
		1 & 0 & 0
	\end{array}\right),\:
	p_5=\left(
	\begin{array}{ccc}
		0 & 0 & 1 \\
		1 & 0 & 0 \\
		0 & 1 & 0
	\end{array}
	\right),\: p_6=\left(
	\begin{array}{ccc}
		0 & 1 & 0 \\
		0 & 0 & 1 \\
		1 & 0 & 0
	\end{array}
	\right).
\end{equation}	

These 6 matrices, $S_{1}$ to $S_{6}$, have been placed in class I of Table 1 and for further discussion would be referred as I$_a$, I$_b$, etc.. Further, interestingly, the 6 structures, i.e., $S_{10}$, $S_{11}$, $S_{12}$, $S_{13}$, $S_{14}$ and $S_{15}$ are also related through permutations and have been placed in class II of the table and would henceforth be referred as II$_a$, II$_b$, etc.. The remaining structure $S_{19 }$ would be discussed separately.  Therefore, instead of discussing 169 possible texture 6 zero combinations, we would be first discussing 144 possibilities of hermitian mass matrices which can be arrived at by considering $M_{U}$ and $M_{D}$ to be from class I and/or class II of Table (\ref{table1}).

\begin{table}
\begin{center}
\begin{tabular}{|c|c|c|}
	\hline 
	& \text{Class I} & \text{Class II} \\ 
	\hline 
a	& $\left(
\begin{array}{ccc}
	0 & A e^{\text{i$\alpha $}} & 0 \\
	A e^{-\text{i$\alpha $}} & 0 & B e^{\text{i$\beta $}} \\ 
	0 & B e^{-\text{i$\beta $}} & C
\end{array}
\right)$ & $\left(
\begin{array}{ccc}
	0 & A e^{\text{i$\alpha $}} & 0 \\
	A e^{-\text{i$\alpha $}} & D & 0 \\
	0 & 0 & C
\end{array}
\right)$ \\ 
\hline 
b	& $\left(
\begin{array}{ccc}
0 & 0 & A e^{\text{i$\alpha $}} \\
0 & C & B e^{-\text{i$\beta $}} \\
A e^{-\text{i$\alpha $}} & B e^{\text{i$\beta $}} & 0
\end{array}
\right)$ & $\left(
\begin{array}{ccc}
0 & 0 & A e^{\text{i$\alpha $}} \\
0 & C & 0 \\
A e^{-\text{i$\alpha $}} & 0 & D
\end{array}
\right)$ \\ 
	\hline 
c	& $\left(
\begin{array}{ccc}
0 & A e^{-\text{i$\alpha $}} & B e^{\text{i$\beta $}} \\
A e^{\text{i$\alpha $}} & 0 & 0 \\
B e^{-\text{i$\beta $}} & 0 & C
\end{array}
\right)$ & $\left(
\begin{array}{ccc}
D & A e^{-\text{i$\alpha $}} & 0 \\
A e^{\text{i$\alpha $}} & 0 & 0 \\
0 & 0 & C
\end{array}
\right)$ \\ 
	\hline 
d	& $\left(
\begin{array}{ccc}
C & B e^{-\text{i$\beta $}} & 0 \\
B e^{\text{i$\beta $}} & 0 & A e^{-\text{i$\alpha $}} \\
0 & A e^{\text{i$\alpha $}} & 0
\end{array}
\right)$ &   $\left(
\begin{array}{ccc}
C & 0 & 0 \\
0 & D & A e^{-\text{i$\alpha $}} \\
0 & A e^{\text{i$\alpha $}} & 0
\end{array}
\right)$  \\ 
	\hline 
e	& $\left(
\begin{array}{ccc}
0 & B e^{\text{i$\beta $}} & A e^{-\text{i$\alpha $}} \\
B e^{-\text{i$\beta $}} & C & 0 \\
A e^{\text{i$\alpha $}} & 0 & 0
\end{array}
\right)$  & $\left(
\begin{array}{ccc}
D & 0 & A e^{-\text{i$\alpha $}} \\
0 & C & 0 \\
A e^{\text{i$\alpha $}} & 0 & 0
\end{array}
\right)$ \\ 
	\hline 
f	& $\left(
\begin{array}{ccc}
C & 0 & B e^{-\text{i$\beta $}} \\
0 & 0 & A e^{\text{i$\alpha $}} \\
B e^{\text{i$\beta $}} & A e^{-\text{i$\alpha $}} & 0
\end{array}
\right) $ & $\left(
\begin{array}{ccc}
C & 0 & 0 \\
0 & 0 & A e^{\text{i$\alpha $}} \\
0 & A e^{-\text{i$\alpha $}} & D
\end{array}
\right)$ \\ 
	\hline 
\end{tabular} 
\end{center}
\caption{Possible texture 3 zero mass matrices belonging to class I and II} 
\label{table1}
\end{table}

Coming to the methodology, it essentially involves considering a possible texture 6 zero combination, i.e., $M_{U}$ and $M_{D}$ being either of the above listed patterns. The viability of the considered combination is ensured by examining the compatibility of the CKM matrix, constructed from a given combination of mass matrices, with the recent one given by Particle Data Group (PDG) \cite{pdg2018}. To this end, as a first step, hermitian matrix $M_{i}$ ($i=U, D$) can be
expressed as
  \begin{equation}     M_{i}=P_{i}^{\dag}M_{i}^{r}P_{i}, \end{equation}
where  $ M_{i}^{r}$ corresponds to the real matrix and $P_{i}$ denotes the phase matrix. The real matrix $M_{i}^{r}$ can then be diagonalized by the orthogonal
transformations $O_{i}$, i.e.,
\begin{equation}
  M_{i}^{diag}= O_{i}^{T}M_{i}^{r}O_{i}=
  O_{i}^{T}P_{i}M_{i}P_{i}^{\dag}O_{i},
  \end{equation}
where $M_{i}^{diag}={\rm diag}(m_{1},-m_{2},m_{3})$, where the subscripts 1, 2 and 3 refer respectively to u, c and t for the up sector and d, s and b for the down sector. In order to examine the viability of the considered combination, one needs to obtain the CKM matrix using the relation
\begin{equation}
V_{CKM}=O_{U}^{T}P_{U}P_{D}^{\dag}O_{D}= V_{U}^{\dag}V_{D},
\end{equation}
where the unitary matrices $V_{U}(= P_{U}^{\dag}O_{U})$ and
$V_{D}(=P_{D}^{\dag}O_{D})$ are the diagonalizing transformations for the
matrices $M_{U}$ and $M_{D}$ respectively. 

To begin with, let us consider the matrix I$_{a}$ of class I, the corresponding  real matrix $ M_{i}^{r}$ can be expressed as 
\begin{equation}
M_{i}^{r}  =\begin{pmatrix}
  0 & |A_{i}| & 0 \\
  |A_{i}| & 0 & |B_{i}| \\
  0 & |B_{i}| & C_{i} \\
\end{pmatrix}
\end{equation}
and $P_{i}$, the phase matrix, is given by
\begin{equation}
    P_{i}=\begin{pmatrix}
      e^{-\text{i$\alpha $}_i} & 0 & 0 \\
      0 & 1 & 0 \\
      0 & 0 & e^{\text{i$\beta $}_i} \\
    \end{pmatrix}.
\end{equation}
It may be mentioned that the rest of the 5 matrices belonging to class I of the table can be similarly expressed in terms of a real matrix $ M_{i}^{r}$ and the corresponding phase matrix $P_{i}$.
An essential step for the construction of the diagonalization transformation is to consider the invariants trace $M_{i}^r$, trace $M_{i}^{r^{2}}$ and determinant $M_{i}^r$ to yield relations involving elements of mass matrices. For all the six matrices belonging to class I of the table, using these invariants, the relations of the matrix elements in terms of quark masses can be expressed as
\begin{equation}
C_{i}=(m_{1}-m_{2}+m_{3}),~~~~|A_{i}|^2+|B_{i}|^2=(m_1m_2+m_2m_3-m_1m_3),~~~~|A_{i}|^2C_{i}=(m_1m_2m_3).
\end{equation}
Corresponding to the matrix I$_{a}$, the diagonalizing transformation $O_{i}$ is given as 
\begin{equation}
O_i\text{  }=\text{     }\left(
\begin{array}{ccc}
	\sqrt{\frac{m_2m_3\left(m_3-m_2\right)}{C_i\left(m_1+m_2\right)\left(m_3-m_1\right)}} & \sqrt{\frac{m_1m_3\left(m_1+m_3\right)}{C_i\left(m_1+m_2\right)\left(m_3+m_2\right)}} & \sqrt{\frac{m_1m_2\left(m_2-m_1\right)}{C_i\left(m_3+m_2\right)\left(m_3-m_1\right)}} \\
	\sqrt{\frac{m_1\left(m_3-m_2\right)}{\left(m_1+m_2\right)\left(m_3-m_1\right)}} & -\sqrt{\frac{m_2\left(m_1+m_3\right)}{\left(m_1+m_2\right)\left(m_3+m_2\right)}} & \sqrt{\frac{m_3\left(m_2-m_1\right)}{\left(m_3-m_1\right)\left(m_2+m_3\right)}} \\
	-\sqrt{\frac{m_1\left(m_1+m_3\right)\left(m_2-m_1\right)}{C_i\left(m_1+m_2\right)\left(m_3-m_1\right)}} & \sqrt{\frac{m_2\left(m_3-m_2\right)\left(m_2-m_1\right)}{C_i\left(m_1+m_2\right)\left(m_3+m_2\right)}} & \sqrt{\frac{m_3\left(m_3-m_2\right)\left(m_3+m_1\right)}{C_i\left(m_3+m_2\right)\left(m_3-m_1\right)}}
\end{array}
\right).	
\end{equation}	
For the other matrices belonging to class I, one can obtain the corresponding diagonalizing transformations $O_{i}$ in a similar manner.

Similarly, for all the matrices belonging to class II of the table, the relations of the mass matrix elements in terms of the quark masses are given by
\begin{equation}
C_{i}+ D_{i}=(m_{1}-m_{2}+m_{3}),~~~~|A_{i}|^2-C_{i}D_{i}=(m_1m_2+m_2m_3-m_1m_3),~~~~|A_{i}|^2C_{i}=(m_1m_2m_3).
\end{equation} For the matrix II$_{a}$, the corresponding  real matrix $ M_{i}^{r}$ can be expressed as
\begin{equation}
M_{i}^{r}=\begin{pmatrix}
  0 & |A_{i}| & 0 \\
  |A_{i}| & D_{i} & 0 \\
  0 & 0 & C_{i} \\
\end{pmatrix},
\end{equation}
with the phase matrix $P_i$ being
\begin{equation}
 P_i=\left(
\begin{array}{ccc}
e^{-\text{i$\alpha $}_i} & 0 & 0 \\
0 & 1 & 0 \\
0 & 0 & 1
\end{array}
\right).
\end{equation}
Further, the corresponding diagonalizing transformation can be expressed as   
\begin{equation}  
O_i=\left(
\begin{array}{ccc}
\sqrt{\frac{m_2}{m_1+m_2}} & \sqrt{\frac{m_1}{m_1+m_2}} & 0 \\
\sqrt{\frac{m_1}{m_1+m_2}} & -\sqrt{\frac{m_2}{m_1+m_2}} & 0 \\
0 & 0 & 1
\end{array}
\right).
\end{equation}
One can obtain similar matrices for the other 5 matrices of class II as well.

As a next step of our analysis, we present all possible texture 6 combinations, wherein, $M_U$ and $M_D$ can be considered from class I and/or class II of the table, leading to the following:
\\
\\
Category 1: $M_{U}$ and $M_{D}$ both from class I. \\
Category 2: $M_{U}$ from class I and $M_{D}$ from class II.\\
Category 3: $M_{D}$ from class I and $M_{U}$ from class II.\\
Category 4: $M_{U}$ and $M_{D}$ both from class II.\\ \\
To begin with, we discuss Category 1 first wherein both the matrices $M_{U}$ and $M_{D}$ can each be any of the 6 possible structures namely I$_{a-f}$. This results into a total of 36 combinations of texture 6 zero mass matrices. Out of the 36 possibilities, we first consider 6 combinations wherein both $M_{U}$ and $M_{D}$ have the same structure, i.e., I$_a$I$_a$, I$_b$I$_b$, etc.. Constructing the corresponding CKM matrices, one finds that all 6 are same, i.e., they have the same expressions for all the 9 CKM matrix elements. These matrix elements are given as follows: 

\begin{eqnarray}	
 V_{ud}=\sqrt{\frac{m_d \left(m_b-m_s\right)}{\left(m_b-m_d\right) \left(m_d+m_s\right)}} \sqrt{\frac{\left(-m_c+m_t\right) m_u}{\left(m_t-m_u\right) \left(m_c+m_u\right)}}~~~~~~~~~~~~~~~~~~~~~~~~~~~~~~~~~~~~~~~~~~~~~~~~~~~~~\nonumber \\
+e^{-\text{i$\phi $}_1} \sqrt{\frac{m_b \left(m_b-m_s\right) m_s}{\left(m_b-m_d\right) \left(m_b+m_d-m_s\right) \left(m_d+m_s\right)}} \sqrt{\frac{m_c m_t \left(-m_c+m_t\right)}{\left(m_t-m_u\right) \left(m_c+m_u\right) \left(-m_c+m_t+m_u\right)}}\nonumber \\+e^{\text{i$\phi $}_2} \sqrt{\frac{m_d \left(m_b+m_d\right) \left(-m_d+m_s\right)}{\left(m_b-m_d\right) \left(m_b+m_d-m_s\right) \left(m_d+m_s\right)}} \sqrt{\frac{\left(m_c-m_u\right) m_u \left(m_t+m_u\right)}{\left(m_t-m_u\right) \left(m_c+m_u\right) \left(-m_c+m_t+m_u\right)}}~
\end{eqnarray}

\begin{eqnarray}
V_{us}= -\sqrt{\frac{\left(m_b+m_d\right) m_s}{\left(m_b+m_s\right) \left(m_d+m_s\right)}} \sqrt{\frac{\left(-m_c+m_t\right) m_u}{\left(m_t-m_u\right) \left(m_c+m_u\right)}}~~~~~~~~~~~~~~~~~~~~~~~~~~~~~~~~~~~~~~~~~~~~~~~~~~~\nonumber \\
+e^{-\text{i$\phi $}_1} \sqrt{\frac{m_b m_d \left(m_b+m_d\right)}{\left(m_b+m_d-m_s\right) \left(m_b+m_s\right) \left(m_d+m_s\right)}} \sqrt{\frac{m_c m_t \left(-m_c+m_t\right)}{\left(m_t-m_u\right) \left(m_c+m_u\right) \left(-m_c+m_t+m_u\right)}}\nonumber \\ 
-e^{\text{i$\phi $}_2} \sqrt{\frac{\left(m_b-m_s\right) m_s \left(-m_d+m_s\right)}{\left(m_b+m_d-m_s\right) \left(m_b+m_s\right) \left(m_d+m_s\right)}} \sqrt{\frac{\left(m_c-m_u\right) m_u \left(m_t+m_u\right)}{\left(m_t-m_u\right) \left(m_c+m_u\right) \left(-m_c+m_t+m_u\right)}}~
\end{eqnarray}

\begin{eqnarray}
V_{ub}=\sqrt{\frac{m_b \left(-m_d+m_s\right)}{\left(m_b-m_d\right) \left(m_b+m_s\right)}} \sqrt{\frac{\left(-m_c+m_t\right) m_u}{\left(m_t-m_u\right) \left(m_c+m_u\right)}}
~~~~~~~~~~~~~~~~~~~~~~~~~~~~~~~~~~~~~~~~~~~~~~~~~~~~~\nonumber \\
+e^{-\text{i$\phi $}_1} \sqrt{\frac{m_d m_s \left(-m_d+m_s\right)}{\left(m_b-m_d\right) \left(m_b+m_d-m_s\right) \left(m_b+m_s\right)}} \sqrt{\frac{m_c m_t \left(-m_c+m_t\right)}{\left(m_t-m_u\right) \left(m_c+m_u\right) \left(-m_c+m_t+m_u\right)}}\nonumber \\ 
-e^{\text{i$\phi $}_2} \sqrt{\frac{m_b \left(m_b+m_d\right) \left(m_b-m_s\right)}{\left(m_b-m_d\right) \left(m_b+m_d-m_s\right) \left(m_b+m_s\right)}} \sqrt{\frac{\left(m_c-m_u\right) m_u \left(m_t+m_u\right)}{\left(m_t-m_u\right) \left(m_c+m_u\right) \left(-m_c+m_t+m_u\right)}}
~
\end{eqnarray}

\begin{eqnarray}
V_{cd}=-\sqrt{\frac{m_d \left(m_b-m_s\right)}{\left(m_b-m_d\right) \left(m_d+m_s\right)}} \sqrt{\frac{m_c \left(m_t+m_u\right)}{\left(m_c+m_t\right) \left(m_c+m_u\right)}}
~~~~~~~~~~~~~~~~~~~~~~~~~~~~~~~~~~~~~~~~~~~~~~~~~~~\nonumber\\
+e^{-\text{i$\phi $}_1} \sqrt{\frac{m_b \left(m_b-m_s\right) m_s}{\left(m_b-m_d\right) \left(m_b+m_d-m_s\right) \left(m_d+m_s\right)}} \sqrt{\frac{m_t m_u \left(m_t+m_u\right)}{\left(m_c+m_t\right) \left(m_c+m_u\right) \left(-m_c+m_t+m_u\right)}}
\nonumber\\
-e^{\text{i$\phi $}_2} \sqrt{\frac{m_d \left(m_b+m_d\right) \left(-m_d+m_s\right)}{\left(m_b-m_d\right) \left(m_b+m_d-m_s\right) \left(m_d+m_s\right)}} \sqrt{\frac{m_c \left(-m_c+m_t\right) \left(m_c-m_u\right)}{\left(m_c+m_t\right) \left(m_c+m_u\right) \left(-m_c+m_t+m_u\right)}}~
\end{eqnarray}

\begin{eqnarray}
V_{cs}=\text{  }\sqrt{\frac{\left(m_b+m_d\right) m_s}{\left(m_b+m_s\right) \left(m_d+m_s\right)}} \sqrt{\frac{m_c \left(m_t+m_u\right)}{\left(m_c+m_t\right) \left(m_c+m_u\right)}}~~~~~~~~~~~~~~~~~~~~~~~~~~~~~~~~~~~~~~~~~~~~~~~~~~~~\nonumber\\
+e^{-\text{i$\phi $}_1} \sqrt{\frac{m_b m_d \left(m_b+m_d\right)}{\left(m_b+m_d-m_s\right) \left(m_b+m_s\right) \left(m_d+m_s\right)}} \sqrt{\frac{m_t m_u \left(m_t+m_u\right)}{\left(m_c+m_t\right) \left(m_c+m_u\right) \left(-m_c+m_t+m_u\right)}}\nonumber\\
+e^{\text{i$\phi $}_2} \sqrt{\frac{\left(m_b-m_s\right) m_s \left(-m_d+m_s\right)}{\left(m_b+m_d-m_s\right) \left(m_b+m_s\right) \left(m_d+m_s\right)}} \sqrt{\frac{m_c \left(-m_c+m_t\right) \left(m_c-m_u\right)}{\left(m_c+m_t\right) \left(m_c+m_u\right) \left(-m_c+m_t+m_u\right)}}
~
\end{eqnarray}

\begin{eqnarray}
V_{cb}= -\sqrt{\frac{m_b \left(-m_d+m_s\right)}{\left(m_b-m_d\right) \left(m_b+m_s\right)}} \sqrt{\frac{m_c \left(m_t+m_u\right)}{\left(m_c+m_t\right) \left(m_c+m_u\right)}}~~~~~~~~~~~~~~~~~~~~~~~~~~~~~~~~~~~~~~~~~~~~~~~~~~~~\nonumber\\
+e^{-\text{i$\phi $}_1} \sqrt{\frac{m_d m_s \left(-m_d+m_s\right)}{\left(m_b-m_d\right) \left(m_b+m_d-m_s\right) \left(m_b+m_s\right)}} \sqrt{\frac{m_t m_u \left(m_t+m_u\right)}{\left(m_c+m_t\right) \left(m_c+m_u\right) \left(-m_c+m_t+m_u\right)}}\nonumber\\
+e^{\text{i$\phi $}_2} \sqrt{\frac{m_b \left(m_b+m_d\right) \left(m_b-m_s\right)}{\left(m_b-m_d\right) \left(m_b+m_d-m_s\right) \left(m_b+m_s\right)}} \sqrt{\frac{m_c \left(-m_c+m_t\right) \left(m_c-m_u\right)}{\left(m_c+m_t\right) \left(m_c+m_u\right) \left(-m_c+m_t+m_u\right)}}~
\end{eqnarray}

\begin{eqnarray}
V_{td}=\sqrt{\frac{m_d \left(m_b-m_s\right)}{\left(m_b-m_d\right) \left(m_d+m_s\right)}} \sqrt{\frac{m_t \left(m_c-m_u\right)}{\left(m_c+m_t\right) \left(m_t-m_u\right)}}~~~~~~~~~~~~~~~~~~~~~~~~~~~~~~~~~~~~~~~~~~~~~~~~~~~~\nonumber\\
+e^{-\text{i$\phi $}_1} \sqrt{\frac{m_b \left(m_b-m_s\right) m_s}{\left(m_b-m_d\right) \left(m_b+m_d-m_s\right) \left(m_d+m_s\right)}} \sqrt{\frac{m_c \left(m_c-m_u\right) m_u}{\left(m_c+m_t\right) \left(m_t-m_u\right) \left(-m_c+m_t+m_u\right)}}\nonumber\\
-e^{\text{i$\phi $}_2} \sqrt{\frac{m_d \left(m_b+m_d\right) \left(-m_d+m_s\right)}{\left(m_b-m_d\right) \left(m_b+m_d-m_s\right) \left(m_d+m_s\right)}} \sqrt{\frac{m_t \left(-m_c+m_t\right) \left(m_t+m_u\right)}{\left(m_c+m_t\right) \left(m_t-m_u\right) \left(-m_c+m_t+m_u\right)}} 
~
\end{eqnarray}

\begin{eqnarray}
V_{ts}= \text{  }-\sqrt{\frac{\left(m_b+m_d\right) m_s}{\left(m_b+m_s\right) \left(m_d+m_s\right)}} \sqrt{\frac{m_t \left(m_c-m_u\right)}{\left(m_c+m_t\right) \left(m_t-m_u\right)}}~~~~~~~~~~~~~~~~~~~~~~~~~~~~~~~~~~~~~~~~~~~~~~~~~~~~\nonumber\\
+e^{-\text{i$\phi $}_1} \sqrt{\frac{m_b m_d \left(m_b+m_d\right)}{\left(m_b+m_d-m_s\right) \left(m_b+m_s\right) \left(m_d+m_s\right)}} \sqrt{\frac{m_c \left(m_c-m_u\right) m_u}{\left(m_c+m_t\right) \left(m_t-m_u\right) \left(-m_c+m_t+m_u\right)}}\nonumber\\
+e^{\text{i$\phi $}_2} \sqrt{\frac{\left(m_b-m_s\right) m_s \left(-m_d+m_s\right)}{\left(m_b+m_d-m_s\right) \left(m_b+m_s\right) \left(m_d+m_s\right)}} \sqrt{\frac{m_t \left(-m_c+m_t\right) \left(m_t+m_u\right)}{\left(m_c+m_t\right) \left(m_t-m_u\right) \left(-m_c+m_t+m_u\right)}}
~
\end{eqnarray}

\begin{eqnarray}
V_{tb}= \sqrt{\frac{m_b \left(-m_d+m_s\right)}{\left(m_b-m_d\right) \left(m_b+m_s\right)}} \sqrt{\frac{m_t \left(m_c-m_u\right)}{\left(m_c+m_t\right) \left(m_t-m_u\right)}}~~~~~~~~~~~~~~~~~~~~~~~~~~~~~~~~~~~~~~~~~~~~~~~~~~~~\nonumber\\
+e^{-\text{i$\phi $}_1} \sqrt{\frac{m_d m_s \left(-m_d+m_s\right)}{\left(m_b-m_d\right) \left(m_b+m_d-m_s\right) \left(m_b+m_s\right)}} \sqrt{\frac{m_c \left(m_c-m_u\right) m_u}{\left(m_c+m_t\right) \left(m_t-m_u\right) \left(-m_c+m_t+m_u\right)}}\nonumber\\
+e^{\text{i$\phi $}_2} \sqrt{\frac{m_b \left(m_b+m_d\right) \left(m_b-m_s\right)}{\left(m_b-m_d\right) \left(m_b+m_d-m_s\right) \left(m_b+m_s\right)}} \sqrt{\frac{m_t \left(-m_c+m_t\right) \left(m_t+m_u\right)}{\left(m_c+m_t\right) \left(m_t-m_u\right) \left(-m_c+m_t+m_u\right)}}
~ ,
\end{eqnarray}
where phases $\phi_{1}=\alpha_{U}-\alpha_{D}$ and
$\phi_{2}=\beta_{U}-\beta_{D}$ are related to the phases
associated with the elements of the mass matrices.

For the purpose of numerical analysis, we first consider the texture combination I$_a$I$_a$ implying $M_{U}$ and $M_{D}$ both being of the form I$_a$. As mentioned earlier, this particular combination corresponds to Fritzsch \emph{ans$\ddot{a}$tz} mentioned in equation (\ref{mumd}). It has been shown \cite{tsmmrrr,tsmmneelu}, without getting into details, that this texture combination is ruled out due to the CKM matrix element $V_{cb}$. As a first step, it would be interesting to present the details regarding the ruling out of this combination, keeping in mind refinements in the measurements of the light quark masses. To this end, we have first investigated the dependence of the matrix element $V_{cb}$ with respect to the light quark masses. As can be see from analytic expressions corresponding to the CKM matrix elements, mentioned above, in order to construct the CKM matrix elements, one needs to provide values of quarks masses as well as phases $\phi_{1}$ and $\phi_{2}$ as inputs. 

The ``current'' quark masses at $M_{Z}$ energy scale \cite{masses} are given by
\begin{eqnarray}
m_{u}=1.45_{-0.45}^{+0.56}~{\rm MeV},~~~m_{d}=2.9_{-0.4}^{+0.5}~{\rm MeV}, 
~~~m_{s}=57.7_{-15.7}^{+16.8}~{\rm MeV},~~~~~~~~~~~  \nonumber 
\\
m_{c}= 0.635 \pm 0.086 ~{\rm GeV},~~~ m_{b}=2.82_{-0.04}^{+0.09}~ {\rm GeV},~~~ m_{t}=172.1\pm 0.6\pm 0.9 ~{\rm GeV}. 
\label{inputs}
\end{eqnarray} 
The most recent lattice values \cite{flag} of the quark mass ratios $\frac{m_u}{m_d}$ and $\frac{m_s}{m_{ud}}$, wherein ${m_{ud}}$ is defined as $\frac{1}{2(m_u + m_d)}$ are 
\begin{equation}
\frac{m_u}{m_d}= 0.45~(3)~~~~~{\rm and}~~~~~\frac{m_s}{m_{ud}}= 27.30~(34).
\label{massrat}
\end{equation}
For the purpose of calculations, in order to investigate to what extent the texture combination I$_a$I$_a$ remains ruled out, we have assumed a relatively wider range of mass $m_u$, i.e, from $0-3.0$ ${\rm MeV}$ and then using the mass ratios mentioned in equation (\ref{massrat}), we have obtained the corresponding wider ranges of masses $m_d$ and $m_s$. Further, in the absence of any information regarding values of the phases $\phi_{1}$ and $\phi_{2}$, these have been given full variation from $0^o$ to $360^o$. Along with these inputs, we have imposed the following recent value as per PDG 2018 \cite{pdg2018} of the precisely known CKM matrix elements $V_{us}$ as a constraint
\begin{equation}
V_{us}= 0.2243 \pm 0.0005.
\end{equation}

Using the above mentioned inputs and constraint, in Figure 1 we have shown the dependence of CKM matrix element $V_{cb}$ with respect to the light quark masses $m_u$, $m_d$ and $m_{s}$. The vertical lines in these plots depict the ranges of these masses given in equation (\ref{inputs}), whereas the narrow experimental range \cite{pdg2018} of the element $V_{cb}$, i.e. $(42.2 \pm 0.8) \times 10^{-3}$, is shown by very closely spaced horizontal lines. From a look at these plots one can note that for the ranges of $m_u$, $m_d$ and $m_{s}$ given in equation (\ref{inputs}), the allowed range of $V_{cb}$ obtained here has no overlap with its experimentally determined range, thereby ruling out this combination of mass matrices. However, interestingly, it appears that in case the lower limits of the light quark masses get pushed slightly lower, the CKM matrix element $V_{cb}$ obtained here would show an overlap with its experimentally determined range. 

\begin{figure}
\begin{multicols}{3}
{\hspace*{-45pt}\includegraphics[width=8cm,height=6cm]{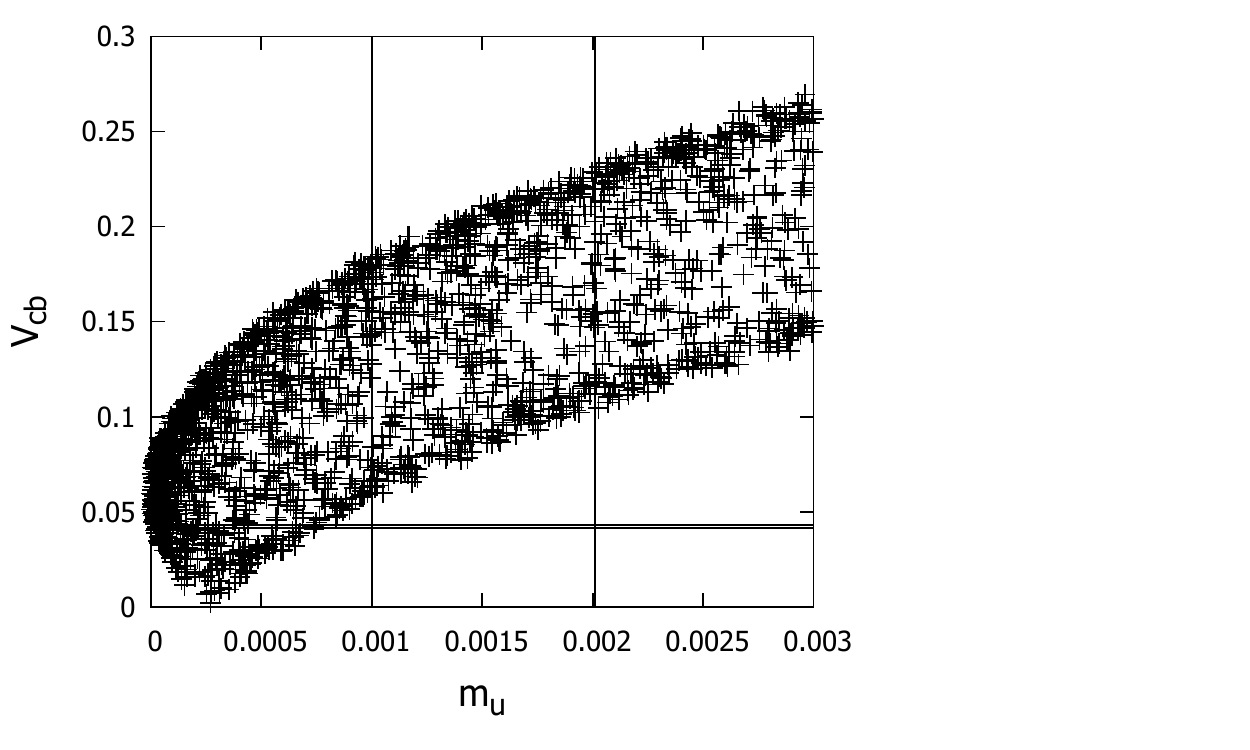}} 
{\hspace*{-8pt}\includegraphics[width=8cm,height=6cm]{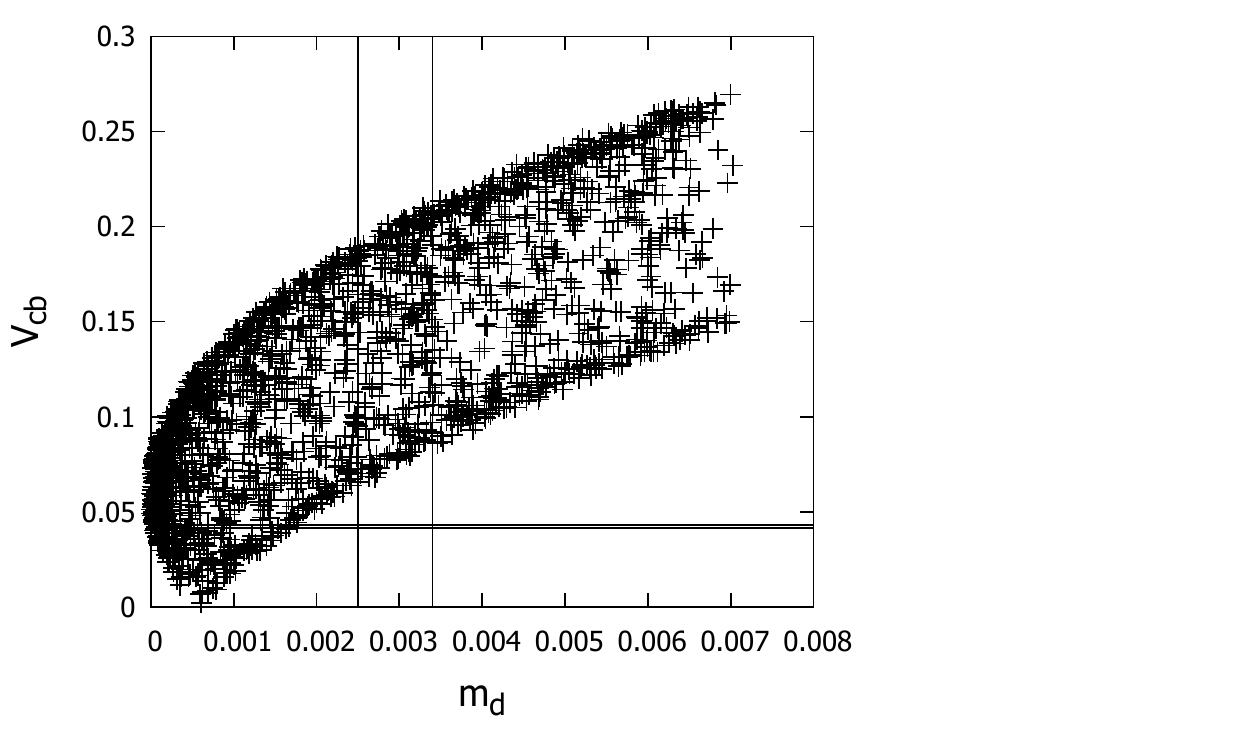}} 
{ \includegraphics[width=8cm,height=6cm]{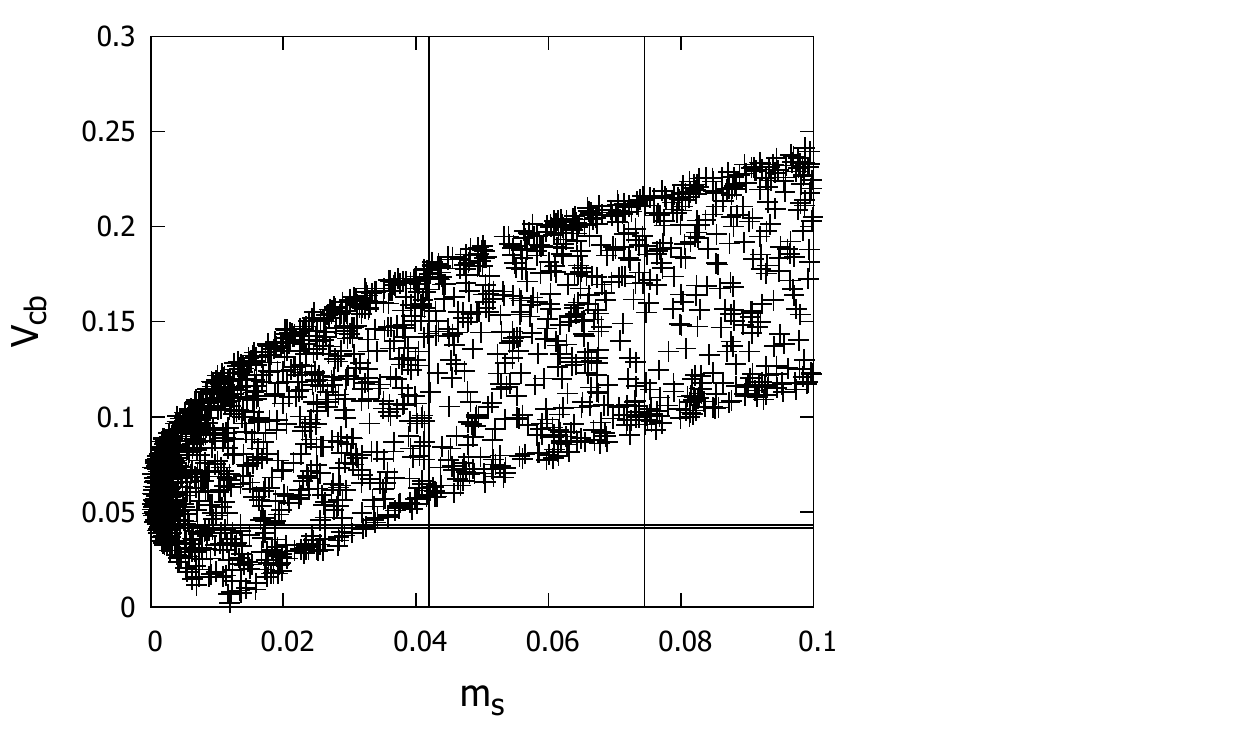}}
\end{multicols} \vspace*{-15pt}
Figure 1: Allowed range of $V_{cb}$ w.r.t the light quark masses obtained by imposing $V_{us}$ as a constraint
\label{figr1}
\end{figure}

To investigate this further, along with  considering the matrix element $V_{us}$ as a constraint, we have also imposed the following value \cite{pdg2018} of the element 
\begin{equation}
V_{ub}= (3.94 \pm 0.36) \times 10^{-3}
\end{equation}
as an another constraint and have again plotted the dependence of $V_{cb}$ on the light quark masses, shown in Figure 2. These plots clearly indicate that when both $V_{us}$ and $V_{ub}$ are imposed as constraints then the allowed range of $V_{cb}$ obtained here lies much outside its experimental range, therefore, completely ruling out the
texture combination I$_a$I$_a$. This conclusion remains valid even if there are considerable changes in the input parameters.

\begin{figure}
\begin{multicols}{3}
{\hspace*{-45pt}\includegraphics[width=8cm,height=6cm]{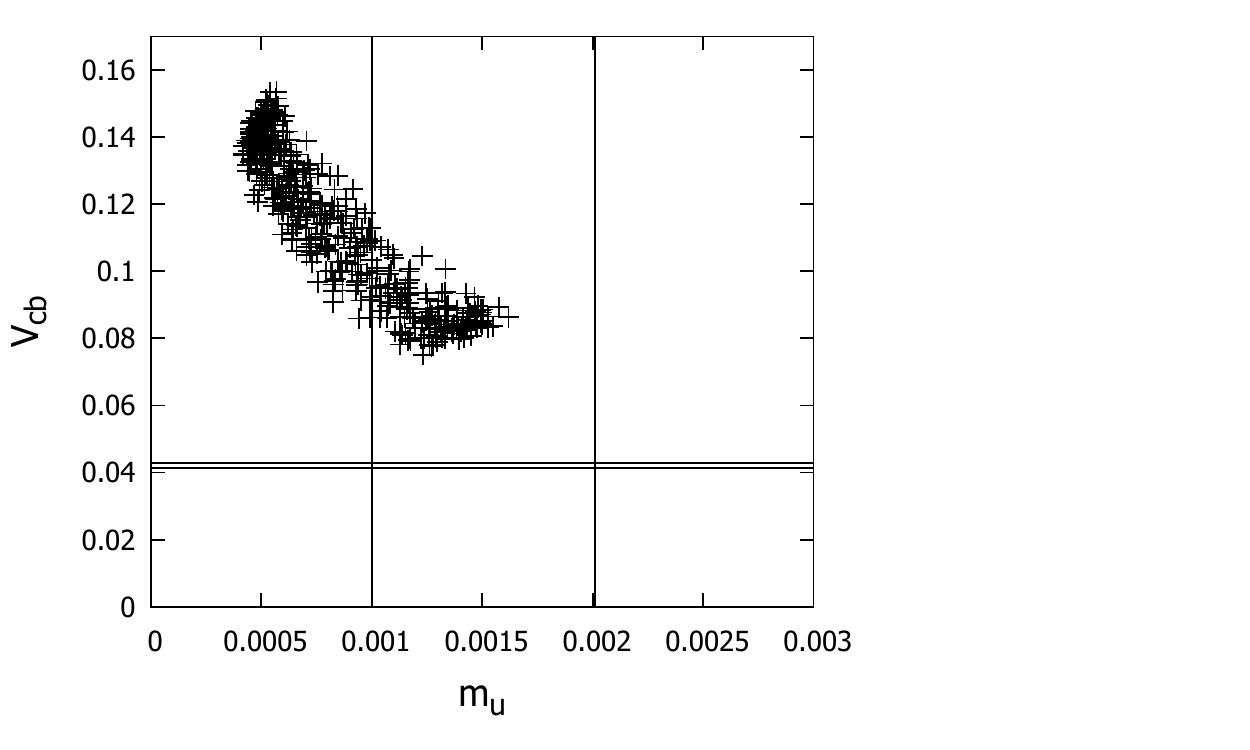}} 
{\hspace*{-8pt}\includegraphics[width=8cm,height=6cm]{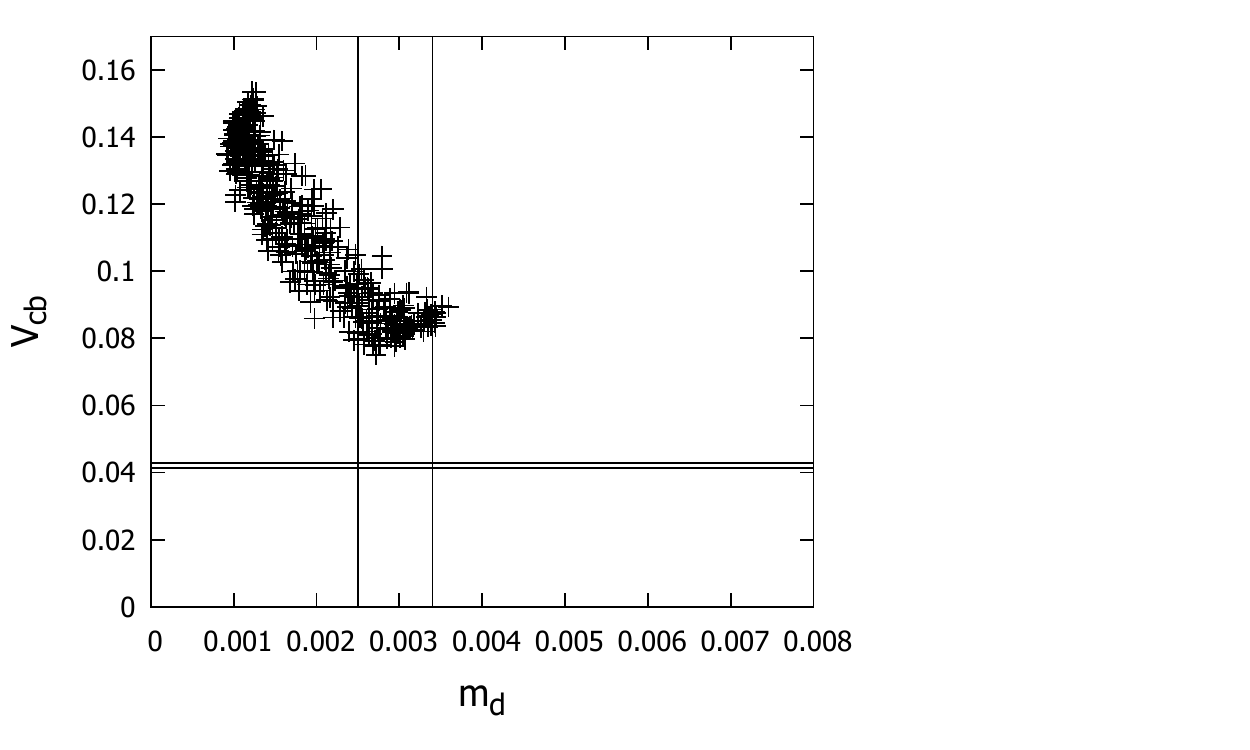}} 
{ \includegraphics[width=8cm,height=6cm]{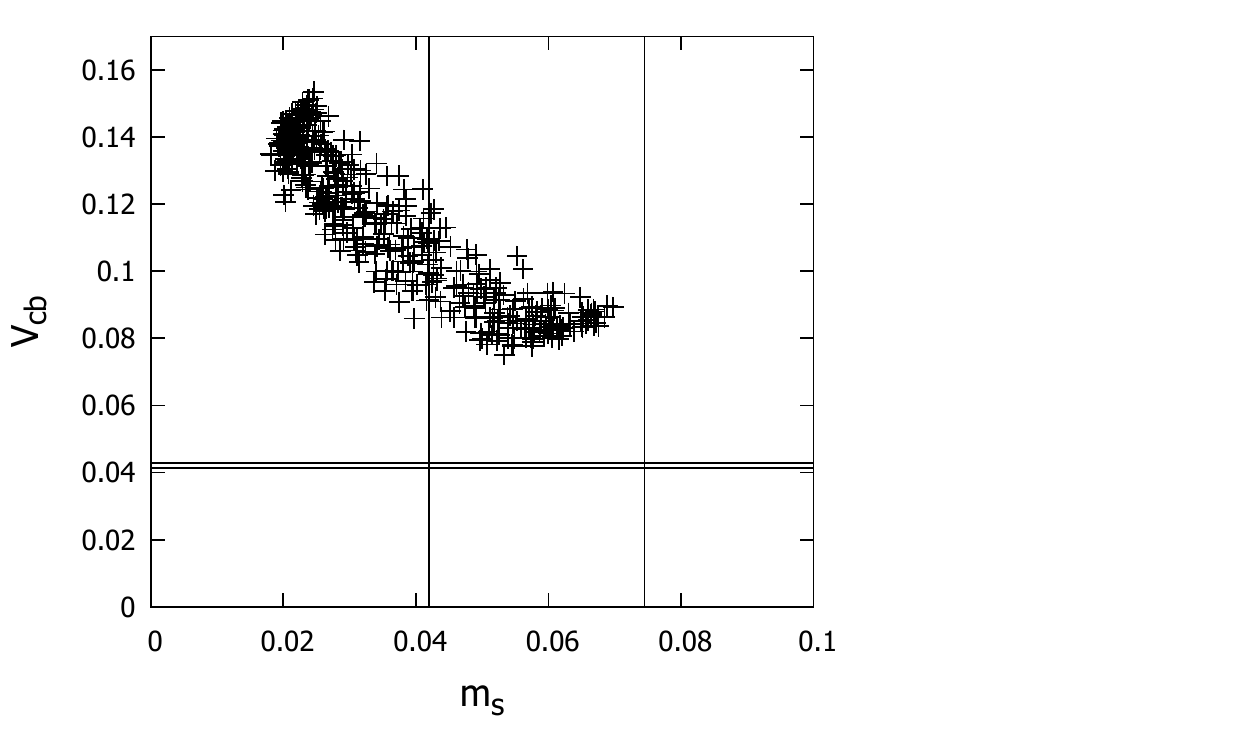}}
\end{multicols} \vspace*{-15pt}
Figure 2: Allowed range of $V_{cb}$ w.r.t the light quark masses obtained by imposing both $V_{us}$ and $V_{ub}$ as constraints 
\label{figr2}
\end{figure}

In order to have a better understanding of the above mentioned results as well as for the sake of completion,  we present the magnitudes of the CKM matrix elements, obtained by considering the masses mentioned in equation (\ref{inputs}) as inputs and both $V_{us}$ and $V_{ub}$ values as constraints, i.e.,
\begin{equation}
V_{CKM}=\begin{pmatrix}
  0.9743-0.9746 & 0.2238-0.2247 & 0.0036-0.0042 \\
  0.2228-0.2241 & 0.9694-0.9718 & 0.0749-0.1018 \\
  0.0168-0.0229 & 0.0731-0.0993 & 0.9947-0.9971 \\
\end{pmatrix}.
\end{equation}
A look at this matrix immediately reveals that the ranges of CKM elements $V_{cb}$, $V_{td}$, $V_{ts}$ and $V_{tb}$ show no overlap with those obtained by recent global analysis as per PDG 2018 \cite{pdg2018}, e.g., 
\begin{equation}
    V_{CKM}=\begin{pmatrix}
      0.9744-0.9746 & 0.2241-0.2250 & 0.0035-0.0038 \\
      0.2239-0.2248 & 0.9735-0.9737 & 0.0414-0.0429 \\
      0.0087-0.0092 & 0.0406-0.0421 & 0.9990-0.9991 \\
    \end{pmatrix}.
\end{equation}
Further, besides determining the quark mixing matrix
elements, we have also evaluated the CP violating phase $\delta$, the Jarlskog's rephasing invariant parameter $J$ and the CP asymmetry parameter $Sin2\beta$, which come out to be 
\begin{equation}
\delta = 79.2^{o}-90^{o},~~~J =(5.84-9.49)\times 10^{-5},~~~Sin2\beta= 0.354-0.430.
 \end{equation}
Again, we find that the above ranges show absolutely no overlap with the experimentally determined ranges \cite{pdg2018} of these quantities given by 
\begin{equation}
\delta = 68.4^{o}-77.7^{o}, ~~~J =(3.03-3.33)\times 10^{-5},~~~Sin2\beta= 0.674-0.708.  
\end{equation}

For the sake of completeness, in Figures 3 and 4 we have presented plots of the Jarlskog's rephasing invariant parameter $J$ as well as the CP asymmetry parameter $Sin2\beta$ respectively with respect to the light quark masses $m_u$, $m_d$ and $m_{s}$. While plotting these graphs, the inputs are the same as for the earlier $V_{cb}$ versus the light quark masses plots, along with both $V_{us}$ and $V_{ub}$ as constraints. Again, these plots reveal that the allowed ranges of these parameters have no overlap with their experimental ranges, presented as solid horizontal lines.

\begin{figure}
\begin{multicols}{3}
{\hspace*{-45pt}\includegraphics[width=8cm,height=6cm]{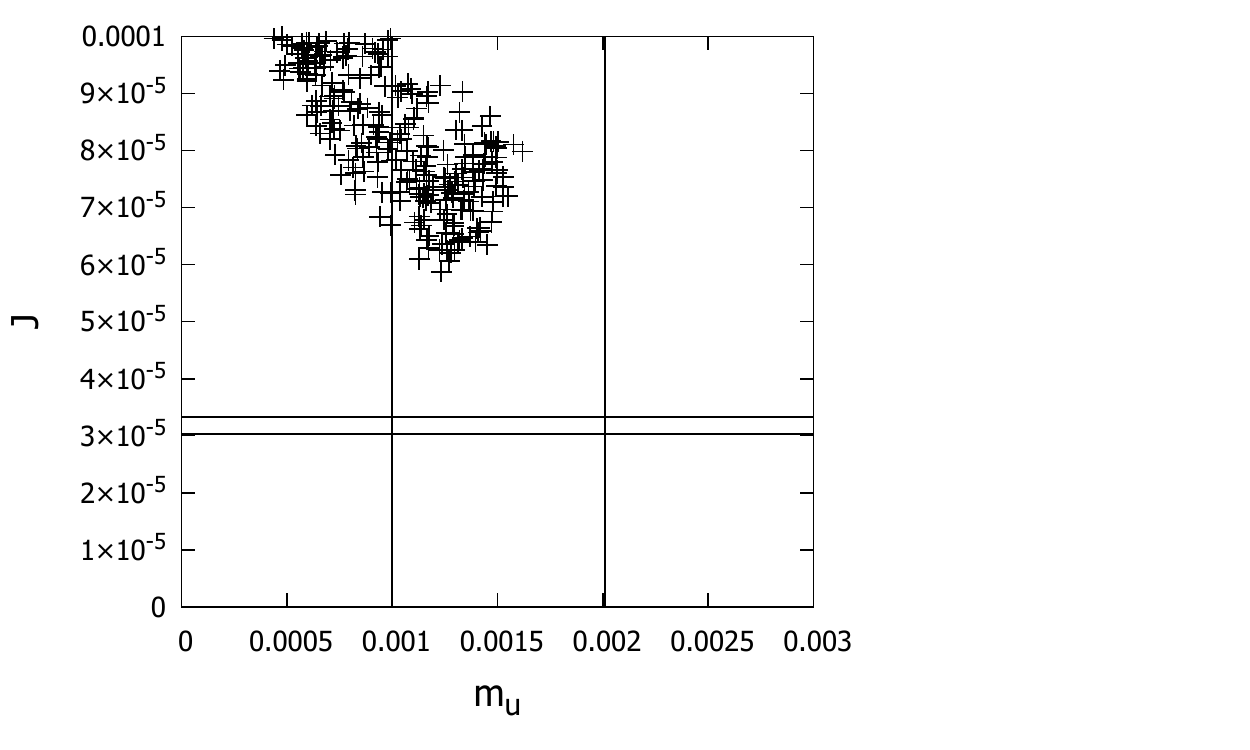}} 
{\hspace*{-8pt}\includegraphics[width=8cm,height=6cm]{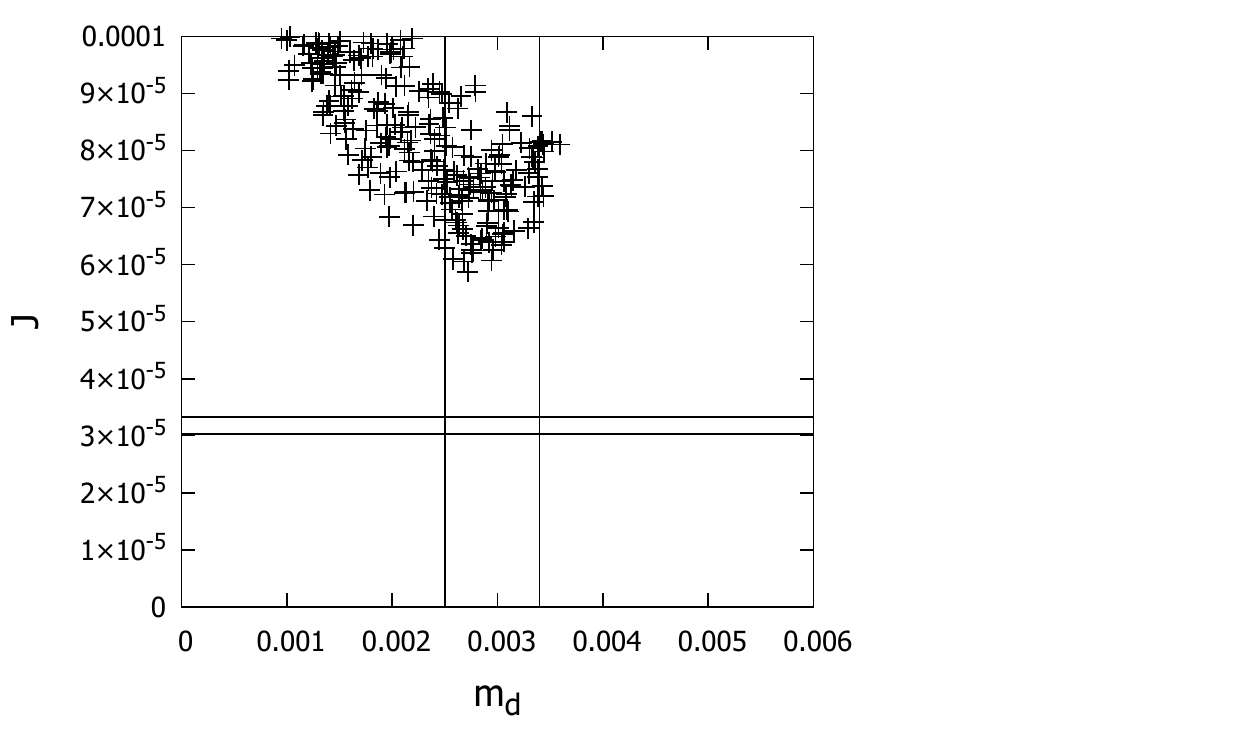}} 
{ \includegraphics[width=8cm,height=6cm]{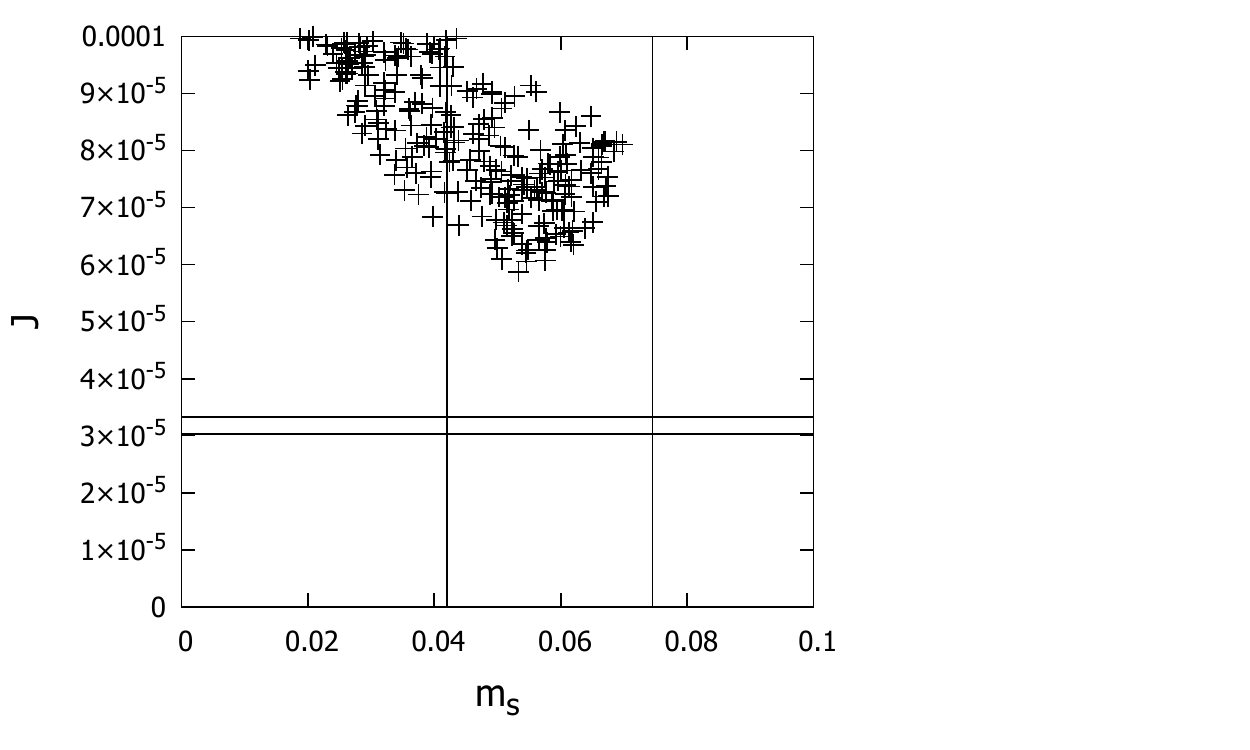}}
\end{multicols} \vspace*{-15pt}
Figure 3:  Allowed range of $J$ w.r.t the light quark masses obtained by imposing both $V_{us}$ and $V_{ub}$ as constraints
\label{figr3}
\end{figure}

\begin{figure}
\begin{multicols}{3}
{\hspace*{-45pt}\includegraphics[width=8cm,height=6cm]{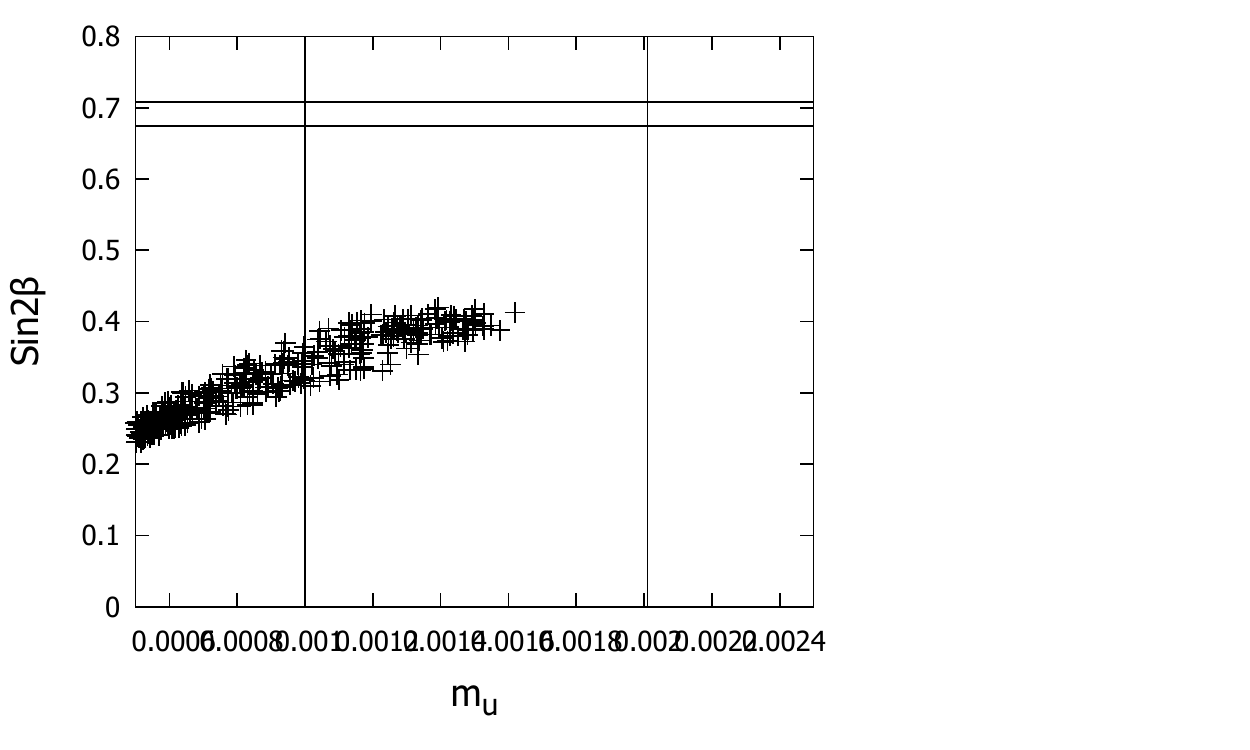}} 
{\hspace*{-8pt}\includegraphics[width=8cm,height=6cm]{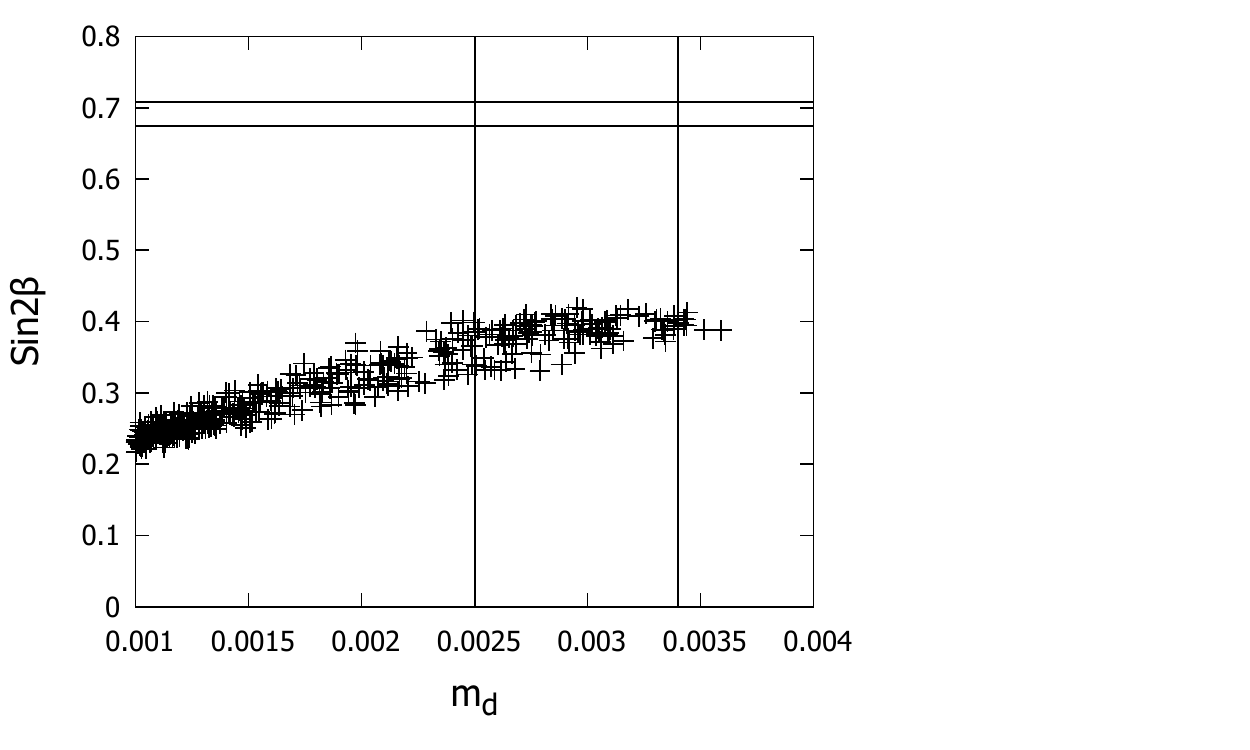}} 
{ \includegraphics[width=8cm,height=6cm]{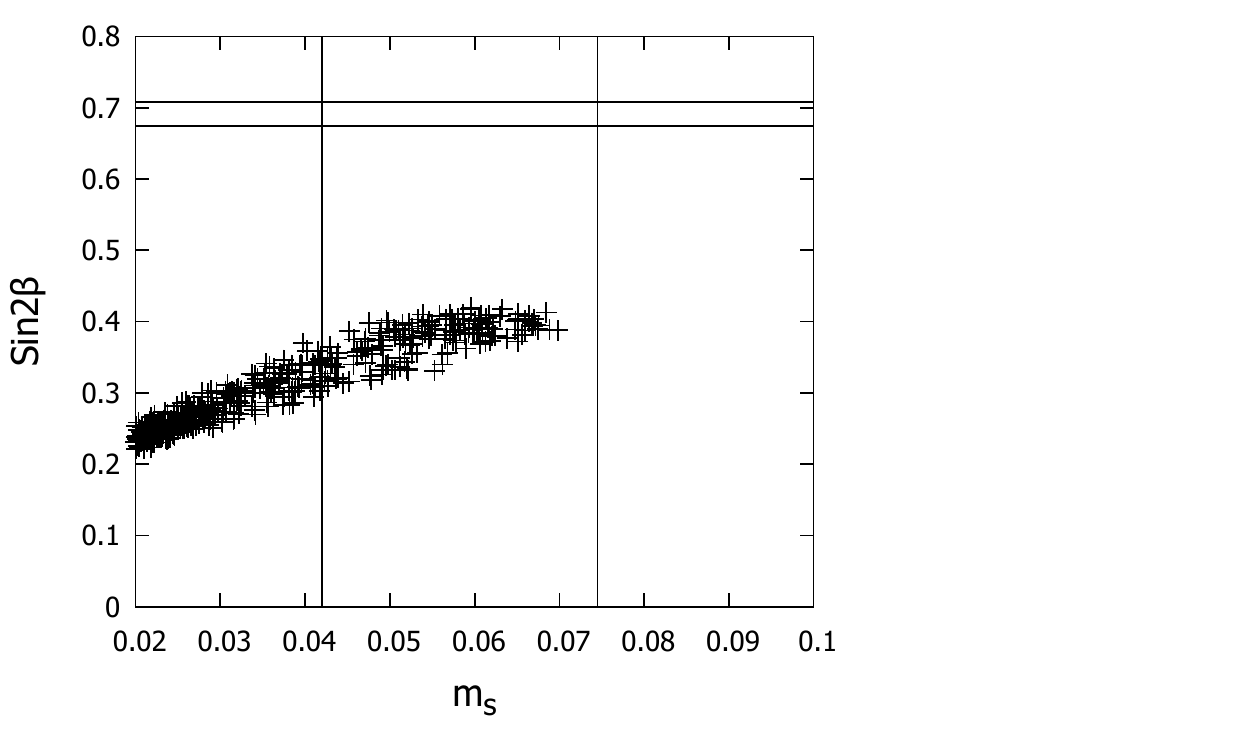}}
\end{multicols} \vspace*{-15pt}
Figure 4: Allowed range of $Sin2\beta$ w.r.t the light quark masses obtained by imposing both $V_{us}$ and $V_{ub}$ as constraints
\label{figr4}
\end{figure}

The above discussion, therefore, leads to the conclusion that not only the texture combination I$_a$I$_a$ gets ruled out, but also, this conclusion remains valid even if there are considerable changes in the input parameters. Similarly, as emphasized earlier, the other such combinations wherein both $M_{U}$ and $M_{D}$ have the same structure are also not compatible with the recent quark mixing data. This can also be checked by the use of the permutation symmetry. 

For the remaining 30 combinations of Category 1, wherein $M_{U}$ and $M_{D}$ can have different structures, i.e., of the type I$_a$I$_b$, I$_c$I$_d$, etc., again the above methodology can be repeated in order to check the viability of the various combinations. Interestingly, for these 30 combinations, one finds that the CKM matrix so obtained does not have the usual structure wherein the diagonal elements are almost unity whereas the off diagonal elements are much smaller than these. For example, considering the combination I$_c$I$_d$, i.e.,  $M_{U}$ having structure I$_c$ and $M_{D}$ having structure I$_d$, the CKM matrix, wherein we have used the hierarchy of quark masses and presented only the leading order terms, so obtained is given by
\begin{equation}
 V_{CKM}=\left(
\begin{array}{ccc}
e^{-\text{i$\alpha $}_U} \sqrt{\frac{m_d}{m_s}} & -e^{-\text{i$\alpha $}_U} & e^{-\text{i$\alpha $}_U} \sqrt{\frac{m_s}{m_b}}+e^{-\text{i$\beta $}_D} \sqrt{\frac{m_u}{m_c}} \\
e^{-\text{i$\beta $}_D} \sqrt{\frac{m_d}{m_b}}+e^{i({\text{$\alpha $}_D+\text{$\beta $}_U})} \sqrt{\frac{m_c}{m_t}} & -e^{-\text{i$\alpha $}_U} \sqrt{\frac{m_u}{m_c}}-e^{-\text{i$\beta $}_D} \sqrt{\frac{m_s}{m_b}} & -e^{-\text{i$\beta $}_D} \\
 e^{{\text{i($\alpha $}_D+\text{$\beta $}_U})} & e^{{\text{i($\alpha $}_D+\text{$\beta $}_U})} \sqrt{\frac{m_d}{m_s}} & e^{-\text{i$\beta $}_D} \sqrt{\frac{m_c}{m_t}}
\end{array}
\right).
\end{equation}
From the above structure of matrix, one can easily find out that off diagonal elements, e.g.,  $V_{us}$, $V_{cb}$ and $V_{td}$ are of the order of unity whereas diagonal elements $V_{ud}$, $V_{cs}$ and $V_{tb}$ are smaller than unity which is in complete contrast to the usual structure of the CKM matix. This can also be seen by carrying out a numerical analysis for this case using the inputs mentioned in equation (\ref{inputs}). The CKM matrix so obtained comes out to be 
\begin{equation}
V_{CKM}=\begin{pmatrix}
  0.2105-0.2234 & 0.9565-0.9723 & 0.0844-0.1973 \\
  0.0167-0.1021 & 0.0743-0.1968 & 0.9784-0.9948 \\
  0.9709-0.9763 & 0.2072-0.2308 & 0.0552-0.0644  \\
\end{pmatrix}.
\end{equation}
The above matrix, clearly, does not have the usual structure of the CKM matrix since the diagonal elements $V_{ud}$, $V_{cs}$ and $V_{tb}$ are much smaller than unity whereas the off diagonal elements $V_{us}$, $V_{cb}$ and $V_{td}$ are approximately 1. Also, this matrix is again not at all compatible with the one given by PDG 2018. Therefore, one can conclude that 30 combinations of Category 1, wherein $M_{U}$ and $M_{D}$ have different structures are also ruled out. It may be interesting to note that even if, in future, there are changes in the ranges of the light quark masses, these 30 combinations would still be ruled out since the structure of the CKM matrices obtained for these are not the usual ones. This, therefore, has implications for models being built using the `top down' approach.

Considering the combinations pertaining to Categories 2 and 3, wherein $M_{U}$ is a matrix from class I of the table and $M_{D}$ is a matrix from class II and vice versa respectively, interestingly, a detailed analyses of all these 36 combinations for each category show results similar to those for Category 1. In particular, in case we consider 6 cases I$_a$II$_a$, I$_b$II$_b$, etc. belonging to Category 2 or 6 cases II$_a$I$_a$, II$_b$I$_b$, etc. pertaining to Category 3, constructing the corresponding CKM matrices, one finds that all 6 are same for each category, i.e., they have the same expressions for all the 9 CKM matrix elements. This can also be checked using the permutation symmetry.  Corresponding to the combination I$_a$II$_a$, the matrix arrived at by using the earlier mentioned inputs is
\begin{equation}
V_{CKM}=\begin{pmatrix}
  0.9744-0.9746 & 0.2238-0.2247 & 0.0025-0.0031 \\
  0.2234-0.2245 & 0.9724-0.9731 & 0.0561-0.0647 \\
  0.0122-0.0143 & 0.0547-0.0632 & 0.9979-0.9984  \\
\end{pmatrix},
\end{equation}
whereas for the case II$_a$I$_a$, we get 
\begin{equation}
V_{CKM}=\begin{pmatrix}
  0.9744-0.9746 & 0.2238-0.2248 & 0.0049-0.0080 \\
  0.2214-0.2230 & 0.9633-0.9670 & 0.1248-0.1508 \\
  0.0282-0.0364 & 0.1218-0.1472 & 0.9885-0.9922  \\
\end{pmatrix}.
\end{equation}
A look at these matrices reveals that although these have the usual CKM matrix structure, i.e., the diagonal elements being nearly unity whereas the off diagonal elements being much smaller, however, one may note that none of these are compatible with the recent one given by PDG 2018, thereby, ruling these out. Similar to Category 1 results, in case one considers the remaining 30 combinations belonging to each category, one finds that the CKM matrices now obtained do not have the usual structure. For example, for Category 2, considering the case I$_c$II$_d$, the leading order CKM matrix obtained is
\begin{equation}
V_{CKM}=\left(
\begin{array}{ccc}
	\sqrt{\frac{m_u}{m_c}} & -e^{-\text{i$\alpha $}_U} \sqrt{\frac{m_s}{m_b}}-e^{\text{i$\alpha $}_D+\text{i$\beta $}_U}\sqrt{\frac{m_u}{m_t}} & e^{-\text{i$\alpha $}_U} \\
	-1 & \sim0 & e^{-\text{i$\alpha $}_U}\sqrt{\frac{m_u}{m_c}} \\
	\sqrt{\frac{m_c}{m_t}} & e^{\text{i$\alpha $}_D+\text{i$\beta $}_U} & e^{\text{i$\alpha $}_D+\text{i$\beta $}_U} \sqrt{\frac{m_s}{m_b}}
\end{array}
\right).
\end{equation}
Similarly, for Category 3, for the combination II$_c$I$_d$, we obtain the following leading order CKM matrix 
\begin{equation}
V_{CKM}=\left(
\begin{array}{ccc}
	e^{-\text{i$\alpha $}_U} \sqrt{\frac{m_d}{m_s}} & -e^{-\text{i$\alpha $}_U} & e^{-\text{i$\alpha $}_U} \sqrt{\frac{m_s}{m_b}}+e^{-\text{i$\beta $}_D} \sqrt{\frac{m_u}{m_c}} \\
	e^{-\text{i$\beta $}_D} \sqrt{\frac{m_d}{m_b}} & -e^{-\text{i$\beta $}_D} \sqrt{\frac{m_s}{m_b}}-e^{-\text{i$\alpha $}_U} \sqrt{\frac{m_u}{m_c}} & -e^{-\text{i$\beta $}_D} \\
	e^{\text{i$\alpha $}_D} & e^{\text{i$\alpha $}_D} \sqrt{\frac{m_d}{m_s}} & \sim0
\end{array}
\right).
\end{equation}	
A look at these two matrices clearly shows that in each of these 1 diagonal matrix element becomes nearly zero indicating these do not have the usual CKM matrix structure, hence ruling out all such possibilities.

Coming to the combinations pertaining to Category 4 wherein both $M_{U}$ and $M_{D}$ are matrices mentioned in class II of the table. A detailed analysis of all these cases shows that unlike the matrices considered in Category 1, the CKM matrices obtained for the 6 possibilities with both the mass matrices having the same structure, i.e., of the type II$_a$II$_a$, II$_b$II$_b$, etc., do not have the usual CKM matrix structure. For example, for the case II$_a$II$_a$, the CKM matrix so obtained is given by
\begin{equation}
\left(
\begin{array}{ccc}
e^{{\text{i($\alpha $}_D-\text{$\alpha $}_U})} \sqrt{\frac{m_s}{m_d+m_s}} \sqrt{\frac{m_c}{m_c+m_u}}+\sqrt{\frac{m_d}{m_d+m_s}} \sqrt{\frac{m_u}{m_c+m_u}}
& e^{{\text{i($\alpha $}_D-\text{$\alpha $}_U})} \sqrt{\frac{m_d}{m_d+m_s}} \sqrt{\frac{m_c}{m_c+m_u}}-\sqrt{\frac{m_s}{m_d+m_s}} \sqrt{\frac{m_u}{m_c+m_u}}
& 0 \\
 -\sqrt{\frac{m_d}{m_d+m_s}} \sqrt{\frac{m_c}{m_c+m_u}}+e^{{\text{i($\alpha $}_D-\text{$\alpha $}_U})} \sqrt{\frac{m_s}{m_d+m_s}} \sqrt{\frac{m_u}{m_c+m_u}}
& \sqrt{\frac{m_s}{m_d+m_s}} \sqrt{\frac{m_c}{m_c+m_u}}+e^{{\text{i($\alpha $}_D-\text{$\alpha $}_U})} \sqrt{\frac{m_d}{m_d+m_s}} \sqrt{\frac{m_u}{m_c+m_u}}
& 0 \\
 0 & 0 & 1 \\
\end{array}
\right)
\end{equation} 

As is evident, the above matrix is clearly ruled out since the elements $V_{ub}$, $V_{cb}$, $V_{td}$ and $V_{ts}$ come out to be 0, whereas the value of the element $V_{tb}$ is 1. The other such 5 combinations also yield similar matrices. Further, all the remaining 30 other combinations with $M_{U}$ and $M_{D}$ not having the same structure result into CKM matrices with the element $V_{tb}$ being 0, thereby ruling out all of these.
\begin{equation}
\left(
\begin{array}{ccc}
 \sqrt{\frac{m_d}{m_d+m_s}} \sqrt{\frac{m_c}{m_c+m_u}} & -\sqrt{\frac{m_s}{m_d+m_s}} \sqrt{\frac{m_c}{m_c+m_u}} &e^{{\text{i($\alpha $}_D-\text{$\alpha $}_U})} \sqrt{\frac{m_u}{m_c+m_u}} \\
 \sqrt{\frac{m_d}{m_d+m_s}} \sqrt{\frac{m_u}{m_c+m_u}} & -\sqrt{\frac{m_s}{m_d+m_s}} \sqrt{\frac{m_u}{m_c+m_u}} & -e^{{\text{i($\alpha $}_D-\text{$\alpha $}_U})} \sqrt{\frac{m_c}{m_c+\text{m}_u}} \\
  \sqrt{\frac{m_s}{m_d+m_s}} &  \sqrt{\frac{m_d}{m_d+m_s}} & 0 \\
\end{array}
\right)
\end{equation}

Finally, we come to the last possibility, wherein one can consider $M_{U}$ and/or $M_{D}$ both having structure $S_{19}$, mentioned earlier while discussing all possible texture 3 zero mass matrices. It is trivial to note that the case wherein both $M_{U}$ and $M_{D}$ have structure $S_{19}$ is ruled out as in this case the corresponding CKM matrix obtained would be a unit matrix. Next, we consider $M_{U}$ being $S_{19}$, whereas $M_{D}$ belongs to Class I. Pertaining to this combination, out of the 6 possible cases, if $M_{D}$ is considered to have structure I{$_a$}, one arrives at CKM matrix having the usual structure, e.g.,
\begin{equation}
V_{CKM}=\begin{pmatrix}
  0.9748-0.9763 & 0.2160-0.2227 & 0.0004-0.0008 \\
  0.2134-0.2209 & 0.9644-0.9682 & 0.1243-0.1518 \\
  0.0281-0.0336 & 0.1211-0.1480 & 0.9884-0.9922  \\
\end{pmatrix},
\end{equation}
this matrix, however, being ruled out by comparing with the one given by PDG 2018. For the remaining 5 cases, the structure of the CKM matrix is not the usual one, hence ruling these out. Further, one can also consider $M_{D}$ having the form $S_{19}$, whereas $M_{U}$ can have any of the 6 structures belonging to Class I. Out of all the 6 CKM matrices obtained pertaining to these combinations, only 1 case, wherein $M_{U}$ has the structure I$_a$ yields CKM matrix having the usual structure, i.e., 
\begin{equation}
V_{CKM}=\begin{pmatrix}
  0.9985-0.9992 & 0.0387-0.0537 & 0.0024-0.0031 \\
  0.0388-0.0539 & 0.9967-0.9974 & 0.0560-0.0647 \\
  0.0000-0.0001 & 0.0561-0.0647 & 0.9979-0.9984  \\
\end{pmatrix},
\end{equation} 
these matrix elements not lying within the range given by PDG 2018, hence this also being ruled out. For the other 5 cases, the CKM matrices arrived at do not have the usual structure.

One can also consider the possibilities wherein either $M_{U}$ or $M_{D}$ has structure $S_{19}$, whereas correspondingly $M_{D}$ or $M_{U}$ respectively belongs to Class II.It can be easily checked that for all such 12 cases, the CKM matrices thus constructed are found to have 4 vanishing elements, hence, ruling out all these possibilities.
  
To summarize, in view of good deal of refinements in the measurements of small quark masses $m_u$, $m_d$ and $m_{s}$ as well as in the CKM matrix elements, we have carried out an extensive analysis of all possible quark mass matrices having minimal texture, implying texture 6 zero quark mass matrices. In all, we have examined 169 possible texture 6 zero combinations, interestingly, many of these combinations can be ruled out analytically. For the remaining, corresponding CKM matrices have been constructed and compared with the latest mixing data. Again, one finds that all these possibilities are excluded by the present quark mixing data. These conclusions remain valid even if, in future, there are changes in the ranges of the light quark masses or if there are small perturbations in the structures of these texture 6 zero mass matrices. In conclusion, all the 169 possible quark mass matrices with minimal texture are ruled out in the present era of precision measurements, having important implications for model building.
\\
\\
{\bf Acknowledgements} \\ The authors would like to thank Chairman, Department of Physics, P.U., for providing facilities to work.

\end{document}